# Low emittance upgrade for CANDLE project

G. S. Zanyan

CANDLE SRI Foundation, Yerevan, Armenia

**Abstract**

To improve the performance of CANDLE [1] synchrotron light source and stay competitive with recently proposed low emittance upgrade programs in the world we have developed new low emittance lattices for CANDLE booster and storage ring. These lattices have been designed taking into account the new developments in magnet fabrication technology and the multi-bend achromat concept. The main design considerations, the linear and non-linear beam dynamics aspects of the modified lattices are presented.

## 1. Introduction

Recently, several new projects with low emittance storage rings were put under realization (MAX-4 [2, 3], Sirius [4]) as well as many existing accelerator centers (DIAMOND, ESRF, SLS, APS, ELETTRA and etc.)[5,6] started the development of low emittance upgrades for their storage rings, aiming to reduce beam emittance by 1-2 orders of magnitude (up to sub nm range) and enable synchrotron radiation experiments in the diffraction limited regime.

The main applied approaches of obtaining low emittance beams are the use of Multi Bend Achromat (MBA) type lattice cells, bending magnets with longitudinal gradient fields and damping wigglers.

Based on the experience of many accelerator centers and achievements in magnet technologies new low emittance lattices for the booster and storage ring of CANDLE light source were developed. This paper summarizes the latest modifications of booster and storage ring lattices, including the nonlinear beam dynamics aspects.

As a first step, new low emittance full energy booster has been designed [7] in order to provide a reliable beam dynamics during the injection and facility top-up operation. The booster new lattice was designed using compact combine-function magnets. It has the same FODO cell structure as the old booster, but the beam emittance is reduced by factor of 4.

As a next step, the study of low emittance lattice for CANDLE [8] storage ring was performed. It was attempted to reduce beam emittance by using more compact magnets with combined fields and keeping the type of cells and the length of the ring circumference unchanged. By this approach it was possible to reduce the beam emittance from 8.4 nm design value to 5.2 nm still providing sufficient dynamic and momentum acceptances.

For further emittance reduction we have studied the option of using MBA lattices, based on their successful realization at several light sources. As a solution, four-bend achromat (4BA) lattice was proposed, which allows reducing the beam emittance up to 1 nm.

The paper is composed as follows: in section 2 the low emittance booster design for CANDLE project with linear and nonlinear beam dynamics is presented. In section 3 main approaches for low emittance storage ring design for CANDLE light source are discussed. Also main parameters and results of linear and nonlinear beam dynamics study are presented. Finally, the summary is given in section 4.

## 2. Low Emittance booster

The injection system of CANDLE storage ring [1] is 100 MeV linac and 3GeV full energy booster. The design lattice of booster consists of four-fold symmetry periods with 2.6m long straight sections between them. Two of the straight sections are occupied by RF-cavities and other straight sections are required for beam injection and extraction. The main parameters of CANDLE design booster presented in table 1.

**Table 1**: Main parameters of the CANDLE booster

| Parameter | Value |
|---|---|
| Injection Energy (MeV) | 100 |
| Extraction Energy (GeV) | 3 |
| Emittance (nm rad) | 74.9 |
| Circumference (m) | 192 |
| Lattice type | FODO |
| Number of periods | 4 |
| Straight section length (m) | 2.6 |
| Max. betafunction (hor./vert.) (m) | 8.5/ 12.2 |
| Natural chrom. (hor./vert.) | -11.19/ -8.63 |
| Betatron tunes (hor./vert.) | 7.6/7.4 |

For the continuous injection into the storage ring it is necessary to have small emittance in booster at full energy. The lattice of CANDLE booster provides 75nm-rad horizontal emittance which makes impossible the top-up injection. One of the main tasks of low emittance upgrade was to design a new low emittance booster lattice, which would provide clean and continuous injection to storage ring.

The description of new booster system with 192 m circumference is given in Ref. [7]. The new lattice consists of four symmetric arcs separated by 5m long straight sections. Each arc has two matching cells and 9 FODO cells, consisting of two combined-function magnets (with integrated quadrupole and sextupole components) separated by 0.55 m long drift (Figure 1 a)).

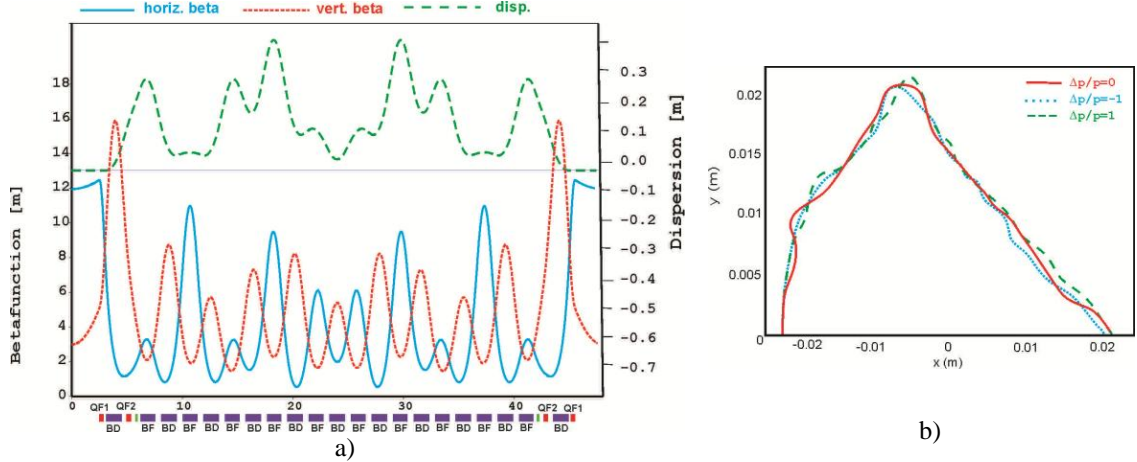

**Figure 1**. a) Beta functions and dispersion in one quadrant of low emittance booster,
b) on and off momentum dynamic apertures

The straight sections provide zero dispersion which is required for the injection and RF systems. For the correction of transverse chromaticities and provision of sufficient dynamic aperture additional defocusing sextupoles are installed in the end of each FODO cell. On and off momentum dynamic apertures at the center of straight section are shown in Figure 1 b).

The lattice with mentioned modifications provides about 4 times lower emittance compared with the original booster lattice. Main parameters of new booster are presented in Table 2.

**Table 2**: Main parameters of the new booster for CANDLE

| Parameter | Value |
|---|---|
| Injection Energy (MeV) | 100 |
| Extraction Energy (GeV) | 3 |
| Emittance (nm rad) | 19.54 |
| Circumference (m) | 192 |
| Lattice type | FODO |
| Number of periods | 4 |
| Straight section length (m) | 5 |
| Max. betafunction (hor./vert.) (m) | 12.4/ 16.1 |
| Natural chrom. (hor./vert.) | -19/ -14.2 |
| Betatron tunes (hor./vert.) | 13.44/8.35 |

## 3. Low emittance storage ring for CANDLE project

The design lattice of CANDLE storage ring whit 216 m circumferences provides 8.4 nm rad beam emittance. It consist of 16 Double-Bend Achromatic (DBA) cells with 4.8 m long straight sections for installation of insertion devices, injection system and RF cavities. In Table 3 the main parameters of the CANDLE storage ring are presented.

**Table 3**: Main parameters of the CANDLE storage ring

| Parameter | Value |
| --- | --- |
| Energy (GeV) | 3 |
| Emittance (nm rad) | 8.458 |
| Circumference (m) | 216 |
| Lattice type | DBA |
| Number of periods | 16 |
| Straight section length (m) | 4.8 |
| Max. betafunction (hor./vert.) (m) | 8.6/17.02 |
| Natural chrom. (hor./vert.) | -18.91/ -14.86 |
| Betatron tunes (hor./vert.) | 13.2/ 4.26 |

Taking into account the well-known relation

$$\varepsilon = A E^2 N^{-3}$$

where $\varepsilon$ is the beam emittance, $E$ is the beam energy, $N$ is the number of bending magnets in the lattice and A is a coefficient defined by the lattice, two approaches of beam emittance reduction were considered.

At first it was attempted to change the original lattice by considering the usage of more compact magnets with combined fields, thus increasing the number of magnets without changing the circumference of the machine and the DBA type of cells [8]. The proposed new lattice contains 24 DBA cells, consisting of two combined function bending and two quadrupole magnets (Figure 2 a)). This new design of storage ring lattice allows reducing the beam emittance up to 5.19 nm rad. The main parameters of proposed lattice are given in Table 4.

**Table 4**. The main parameters of low emittance DBA lattice

| Parameter | Value |
| --- | --- |
| Energy (GeV) | 3 |
| Emittance (nm rad) | 5.19 |
| Circumference (m) | 216 |
| Lattice type | DBA |
| Number of periods | 24 |
| Straight section length (m) | 4.4 |
| Max. betafunction (hor./vert.) (m) | 5/22 |
| Natural chrom. (hor./vert.) | -13.64/ -24.27 |
| Betatron tunes (hor./vert.) | 14.17/3.19 |

In Figure 2 b) on and off momentum dynamic apertures at the center of straight section are shown.

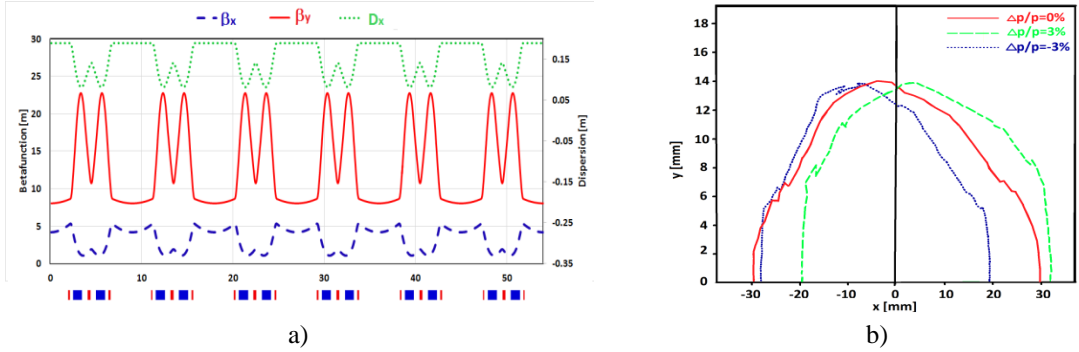

**Figure 2**. a) Beta functions and dispersion in one period of low emittance DBA lattice, b) on and off momentum dynamic apertures at the center of straight section of low emittance DBA lattice

Next, we have studied the possibility of further emittance reduction using MBA lattices. Several types of MBA lattices with different numbers of bending magnets per cell were considered. It was aimed to keep the number of straight sections not less than it is in original lattice. Finally, the search ended up on 4BA cell solution, which with moderate values of magnet parameters provides about 1 nm beam emittance. The new lattice is composed of 16 4BA cells separated by 4 m long straight sections. In each cell the horizontal focusing is performed by 7 quadrupole magnets and the vertical focusing - by 2 quadrupoles and 4 combined-function bending magnets (Figure 3 a)). For chromaticity correction sextupole components of bending magnets and one family of sextupole magnets are used. On and off momentum dynamic apertures at the center of straight section are shown in Figure 3 b). The main parameters of 4BA lattice are given in Table 5.

Note that, the ring circumference for this solution, however, is longer by 44 m compared with original CANDLE lattice, which is the price to pay for beam emittance reduction and keeping the number of straight sections unchanged.

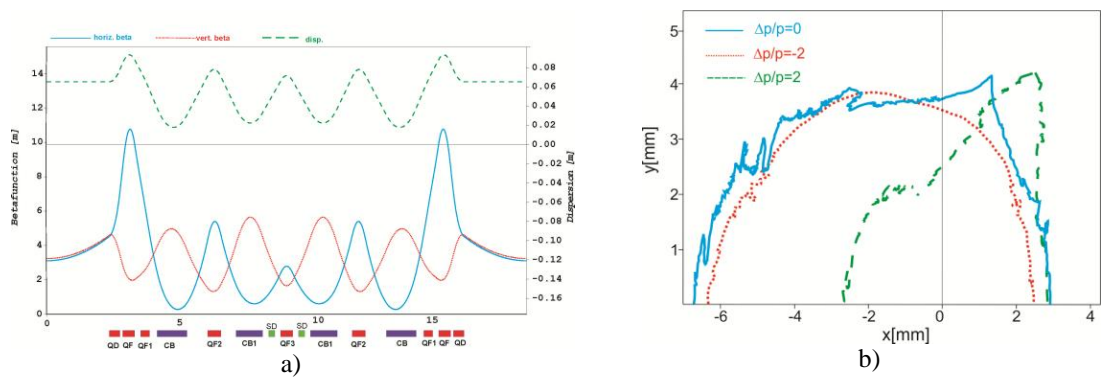

**Figure 3**. a) beta functions and dispersion in one period of low emittance 4BA lattice, b) on and off momentum dynamic apertures at the center of straight section of low emittance 4BA lattice

Table 5. The main parameters of 4BA lattice

| Parameter | Value |
|---|---|
| Energy (GeV) | 3 |
| Emittance (nm rad) | 1.1 |
| Circumference (m) | 257.28 |
| Lattice type | 4BA |
| Number of periods | 16 |
| Straight section length (m) | 4.2 |
| Max. betafunction (hor./vert.) (m) | 10.6/6 |
| Natural chrom. (hor./vert.) | -46.4/ -19.7 |
| Betatron tunes (hor./vert.) | 25.36/13.57 |

## 4. Summary

This paper summarizes the results of study on development of new low emittance lattice options for CANDLE light source booster and storage ring.

New design of booster lattice based on implementation of combined function bending magnets is proposed, which allows reducing the beam emittance at full energy by factor of 4 compared with original booster lattice, granting a reliable beam dynamics during the injection and facility top-up operation.

Two options of low emittance lattices for CANDLE storage ring were considered. In the first solution the emittance was reduced by increasing the number of bending magnets in the lattice keeping the circumference of the ring and the type of lattice unchanged. This could be done by using compact combined-function magnets taking into account the recent achievements in magnet technologies.

As a second option the implementation of MBA lattices was studied. 4BA solution was proposed which allows reducing the beam emittance up to 1 nm. However, as it was aimed to maintain the number of straight sections not less than it is in original lattice, the ring circumference for this solution became 257m (by 41m longer than for original lattice). Also, in comparison with original lattice the proposed lattice is characterized by lowered dynamic aperture, which necessitates the development of more precise injection schemes.

## Acknowledgments

This work was supported by State Committee of Science MES RA, in frame of the research project № SCS 13YR-1C0007.

# References

[1] V. Tsakanov et al., *CANDLE Design report*, 2002.

[2] S.C. Leemann et al., *Status of the MAX IV storage rings*, Proceedings of IPAC'10, Kyoto, Japan, [WEPEA058].

[3] M. Eriksson et al., *The MAX IV synchrotron light source*, Proceedings of IPAC2011, San Sebastián, Spain, [THPC058].

[4] L. Liu et al., *update on SIRIUS, the new Brazilian light source,* Proceedings of IPAC2014, Dresden, Germany, [MOPRO048].

[5] J-L. Revol, P. Berkvens et al., *ESRF upgrade phase II status,* Proceedings of IPAC2014, Dresden, Germany, [MOPRO055].

[6] A. Streun, *Design studies for an upgrade of the SLS storage ring,* Proceedings of IPAC2015, Richmond, VA, USA, [TUPJE047].

[7] G. Zanyan et al., *Low emittance booster design for CANDLE storage ring,* Proceedings of IPAC2011, San Sebastián, Spain, [HPC137].

[8]  G. Zanyan et al., *Low emittance storage ring design for CANDLE project,* Proceedings of IPAC2014, Dresden, Germany, [MOPRO047].